\documentclass[twocolumn]{aastex631}

\usepackage{amsmath}

\begin{document}

\title{Supermassive black holes stripping a subgiant star down to its helium core: \\a new type of multi-messenger source for LISA}

\correspondingauthor{Aleksandra Olejak}
\email{aolejak@mpa-garching.mpg.de}

\author[0000-0002-6105-6492]{Aleksandra Olejak}
\affiliation{Max Planck Institute for Astrophysics,
Karl-Schwarzschild-Straße 1,
85748 Garching b.~München, Germany}

\author[0000-0003-2340-8140]{Jakob Stegmann}
\affiliation{Max Planck Institute for Astrophysics,
Karl-Schwarzschild-Straße 1,
85748 Garching b.~München, Germany}

\author[0000-0001-9336-2825]{Selma E.~de Mink}
\affiliation{Max Planck Institute for Astrophysics,
Karl-Schwarzschild-Straße 1,
85748 Garching b.~München, Germany}
\affiliation{Ludwig-Maximilians-Universität München, Geschwister-Scholl-Platz 1, 80539 München, Germany}

\author[0000-0003-3456-3349]{Ruggero Valli}
\affiliation{Max Planck Institute for Astrophysics,
Karl-Schwarzschild-Straße 1,
85748 Garching b.~München, Germany}

\author[0000-0002-1084-3656]{Re'em Sari}
\affiliation{Racah Institute of Physics, The Hebrew University of Jerusalem, 9190401, Israel}

\author[0000-0001-7969-1569]{Stephen Justham}
\affiliation{Max Planck Institute for Astrophysics,
Karl-Schwarzschild-Straße 1,
85748 Garching b.~München, Germany}



\begin{abstract}

Some stars orbiting supermassive black holes (SMBH) are expected to undergo a gravitational-wave (GW)-driven inspiral and initiate mass transfer on nearly circular orbits. However, the stability and duration of such phases remain unexplored. In this work, we focus on the evolution of a low-mass, radiative-envelope subgiant star being stripped by an SMBH. We find that such systems can undergo a long-lasting, stable mass-transfer phase, even if none of the angular momentum of the transferred material returns to the orbit to counterbalance the GW-driven decay. We show an example where a $2 M_{\odot}$ subgiant is stripped before entering the LISA band and loses almost its entire hydrogen envelope. The remaining helium core undergoes a prolonged GW-driven inspiral, becoming a loud LISA source. If formed in our galaxy, the system would be detectable for several hundred thousand years, ultimately reaching extreme signal-to-noise ratios of a million. Hydrogen shell flashes in the residual envelope cause temporary radial expansions of the stripped star. As a result, a few additional phases of rapid mass transfer occur at orbital periods of 20~--~30 hours. Eventually, the core possibly undergoes circular partial tidal disruption at an orbital period of $\sim$10 minutes, corresponding to a GW emission frequency of a few mHz. We estimate a chance of about 1\% that such a detectable LISA source exists in our own galactic center. The loud final GW transient may lead to a few detections reaching as far as $\sim$$1 \, \rm Gpc$, including, e.g., the Abell clusters.

\end{abstract}



\section{Introduction} \label{sec:intro}

Galactic centers are dense environments where dynamical interactions between stars, as well as between stars and a supermassive black hole (SMBH), are common. While tidal disruption events (TDEs) are typically associated with stars on highly eccentric orbits \citep{Magorrian1999,Gezari2021}, it has been estimated that approximately once every few million years, a star in a Milky-Way like galaxy begins mass transfer onto an SMBH on a mildly eccentric orbit \citep{LinialSari2023}.

The Laser Interferometer Space Antenna (LISA) \citep{AmaroSeoane2017,Colpi2019,LISA2023}, possibly along with other planned missions, TianQin \citep{Luo2016,ChineseGW2021} or Taiji \citep{Taiji2018,ChineseGW2021}, will open a new observational window for gravitational-wave (GW) astronomy. These missions will fill the GW frequency gap from about $0.1\,\rm mHz$ to $1\,\rm Hz$, enabling the detection of new types of sources that are inaccessible to current and future ground-based instruments. Extreme mass ratio inspirals (EMRIs) are among the most promising sources for LISA, where stellar-mass objects emit GWs while orbiting an SMBH. While most studies focus on EMRIs of compact objects such as black holes \citep[e.g.,][]{Amaro-Seoane2007}, this work investigates a novel and yet not well-explored type of EMRI where the inspiraling object is a star. 

In quiescent types of galaxies, stellar EMRIs are thought to form by two primary mechanisms: single-single scattering between stars or tidal capture via the Hills mechanism \citep{Hills1975}. In active galaxies, formation involves migration and trapping of stellar orbits within disk environments, for example, if two SMBHs merge \citep{dorazio2025}. 

Stars transferring mass onto an SMBH have been suggested as potential fuel for active galactic nuclei \citep{Hameury1994}. Theoretical predictions for stellar EMRIs may explain new types of X-ray transients observed at the centers of galaxies. Recent models suggest that quasi-periodic eruptions (QPEs) \citep{Miniutti2019,Giustini2020,Chakraborty2021,Arcodia2021} could provide indirect evidence for stars in close proximity to SMBHs. Proposed explanations for QPEs include mass transfer from main-sequence stars or stripped stellar cores \citep{Dai2013,King2020,King2022,Krolik2022,Chen2022,Metzger2022,Wang2024,LinialSari2023,Lu2023}. However, recent observations \citep{Nicholl2024,Bykov2024} seem to favor models where QPEs arise from interactions between a star (or stellar-origin compact object) with an accretion disk formed after former TDEs \citep{Xian2021,Franchini2023,Linial2023,Tagawa2023,Yao2024}. The proximity to an SMBH is expected to result in strong GW emission and motivated the first efforts to estimate the GW detectability of observed QPE systems \citep{Dai2013,Chen2022,LinialSari2023,Kejriwal2024}. Due to their GW and electromagnetic (EM) counterparts, stellar EMRIs are considered candidates for multi-messenger astronomy, serving as cosmological sirens \citep{Lyu2025}. 

In this work, we focus on the case of a subgiant star with a radiative envelope. Our analysis is the first to incorporate a detailed stellar structure of an evolved star (as derived from a detailed 1D evolutionary code) for a subgiant in the context of mass transfer onto an SMBH, moving beyond the simplified polytropic models commonly used. By considering the detailed envelope structure, we explore the star’s evolution as it approaches and interacts with an SMBH, emphasizing its potential detectability with LISA.

We track the star through multiple mass-transfer phases with its SMBH companion, demonstrating that it is ultimately stripped down to a hot helium core and characterizing the resulting GW signal. Additionally, we estimate the number of such systems that could be detectable by LISA in the local Universe and assess the likelihood of one existing in the Galactic Center. Our study highlights the potential of stellar EMRIs as multi-messenger sources, offering new insights into mass-transfer processes and gravitational dynamics near SMBHs.

\begin{figure*}[ht!] 
     \begin{center}
        \includegraphics[width=0.38\textwidth]{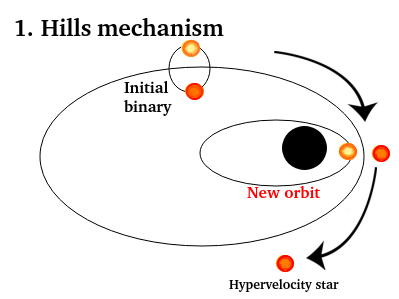}
        \vspace{2ex}
        \includegraphics[width=0.38\textwidth]{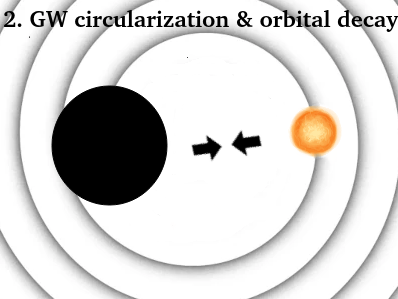}
        \hspace{2ex}\\
        \includegraphics[width=0.38\textwidth]{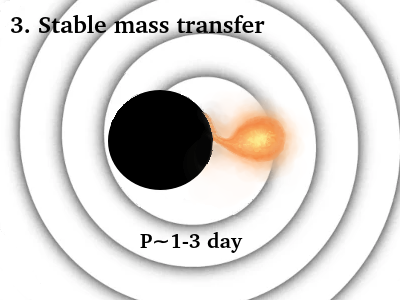}
        \vspace{2ex}
        \includegraphics[width=0.38\textwidth]{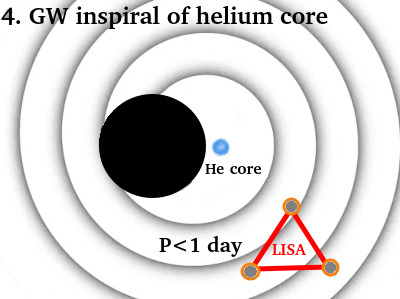}
    \end{center}
    \caption{%
       Cartoon representation of four relevant evolutionary stages of the system. Upper left panel: the initial binary star system enters the Hill sphere of a supermassive black hole (SMBH) and is disrupted. One star becomes gravitationally bound to the SMBH in an eccentric orbit, while the other is ejected from the Galactic nucleus, becoming a hypervelocity star. Upper right panel: the orbit of the star bound to the SMBH gradually shrinks and circularizes due to the strong emission of gravitational waves (GWs). Lower left panel: as the orbit continues to tighten, a subgiant star fills its Roche lobe, initiating a phase of long, stable mass transfer onto the SMBH. Lower right panel: the star is eventually stripped of its hydrogen envelope during the mass transfer process. The system continues to shrink due to GW emission, eventually entering the LISA frequency band as a detectable GW source.
     }%
   \label{fig: Cartoon_stages}
\end{figure*}

\section{Method} \label{sec: Method}

We use the stellar evolution code {\tt MESA} \citep[version r23.05.1,][]{Paxton2011,Paxton2013,Paxton2015,Paxton2018,Paxton2019, Jermyn2023} to study the evolution of a subgiant star with metallicity $Z=0.02$ transferring mass onto an SMBH companion. We opt to study a subgiant star, which did not yet develop a deep convective envelope and possesses unique properties enabling it to undergo stable mass transfer, despite strong GW-driven orbital shrinkage that is not counterbalanced by mass transfer in our models (see Method, Sec.~\ref{subsec: MassTransfer} and ~\ref{subsec: OrbitalEvol}). 

In particular, slightly evolved subgiants initiate mass transfer at wider separations compared to main sequence stars. GW-driven orbital decay strongly depends on the current separation \citep{Peters_1963} and, therefore, is significantly weaker at the onset of mass transfer for a subgiant than for a main-sequence star. This enhances the stability of mass transfer and avoids a runaway process that leads to rapidly increasing mass-transfer rates. Furthermore, such a radiative subgiant star’s entropy profile and the presence of an already-formed helium core allow for an extended GW-driven inspiral phase after the mass transfer ends. These properties make radiative subgiants particularly promising progenitors for long-lived GW sources detectable by LISA.

We choose a donor mass \( M_{*} = 2 \, M_{\odot} \) with a main-sequence lifetime ($\sim 1.5$ Gyr), which allows it to become a subgiant well within the age of the Universe. At the same time, $2~{M_{\odot}}$ stars are still relatively common in stellar populations \citep{Salpeter1955}. Moreover, \citet{Barak2025} found that the TDE rate for moderately massive stars ($M_{\star} \gtrsim 1.0 \, M_{\odot}$) is enhanced relative to their abundance in the stellar population. This enhancement, driven by mass segregation within the sphere of influence, aligns with observational evidence suggesting an over-representation of such moderately massive stars among TDEs \citep{Mockler2022ApJ}.

For the SMBH, we adopt the mass of the Milky Way's Sagittarius A* \( M_{\bullet} = 4.3 \times 10^6 M_{\odot} \) \citep{Ghez2003,Gravity2023}, as the presence of a source in Galactic Center would produce the strongest GW signal. Moreover, the mass of Sagittarius A* aligns with current observations of QPEs \citep{Miniutti2019,Arcodia2021} which are typically associated with SMBHs of relatively low mass (\( M_{\bullet} \approx 10^5-10^7 M_{\odot} \)) and are potentially linked to stellar EMRI. 

\subsection{Mass Transfer} \label{subsec: MassTransfer}

We start our simulations by modeling a subgiant. Once the central hydrogen fraction depletes, we put the star on a circular orbit around an SMBH where it fills its Roche Lobe. The separation $a_{\rm Rlo}$ at which a $2 M_{\odot}$ subgiant star with radius $R_{\star} \approx 4.5 R_{\odot}$ fills its Roche Lobe ${R_{\rm \star}=R_{\rm Rl}}$ on a circular orbit is $a_{\rm Rl} \approx 1200 R_\odot $ \citep{Eggleton1983}. That corresponds to $\approx 65 R_{\rm sch}$ Schwarzschild radii of the SMBH and an orbital period of $\sim 2.6$ day. Mass transfer rates are obtained based on \cite{Kolb1990}, i.e.\ the ``Kolb'' scheme in {\tt MESA}. We assume that the orbit had time to circularize due to GW emission before the mass transfer occurred. However, in reality, the orbit may still be slightly eccentric (see Fig.\ 2 of \citealt{LinialSari2023}). For a discussion on the possible consequences of our simplification, see Section \ref{sec: Discussion}.


The geometry of a star-SMBH system, specifically, the ratio (and the corresponding specific angular momentum) of the binary orbit to the innermost stable circular orbit (ISCO) differs significantly from systems where both components are of stellar origin. This impacts the accretion flow and the evolution of the orbital angular momentum during mass transfer. In particular, it is uncertain whether some part of the angular momentum lost in the stream is transferred back to the orbit via gravitational interaction between the star and an accretion disc if a disc is formed. On the other hand, the star may lose additional orbital angular momentum than is usually assumed during mass transfer, e.g., in case mass leakage occurs through the outer Lagrange point \citep{Linial2017}. These factors have important implications for the stability of mass transfer. In our analysis, we assume that the angular momentum transferred from the star to the SMBH companion is not returned to the system's orbit, following, e.g., \citet{Linial2017} and \citet{Lu2023}. For more discussion of the potential consequences of this assumption, see Sec.~\ref{sec: Discussion}. In order to reflect this approach in our {\tt MESA} simulations, we adopt fully non-conservative mass transfer, in which mass is lost with the specific angular momentum of the donor.

\subsection{Orbital evolution} \label{subsec: OrbitalEvol}

The orbital evolution of a star transferring mass to an SMBH companion is determined by two primary effects: GW emission $\dot{a}_{\rm GW}$ and mass transfer $\dot{a}_{\rm MT}$. So the total orbital evolution is:
$$\dot{a} = \dot{a}_{\rm GW} + \dot{a}_{\rm MT}$$

\noindent where GW emission causes the circular orbit to shrink $\dot{a}_{\text{GW}}$ as \cite{Peters_1963}.

In classical binary star systems, for the relevant mass-ratio regime ($M_{\rm don}/M_{\rm acc} < 1$), conservative mass transfer could efficiently widen the orbit \citep{Tauris2006}, possibly quenching shrinkage by GW emission. However, in the studied star-SMBH type of systems, angular momentum is not expected to be transferred back to the orbit. Following \cite{Lu2023}, we adopt a fully non-conservative model with the mass lost by the donor star carrying away the specific angular momentum of the donor into our MESA model. This approach mimics a real scenario where the SMBH accretes mass without transferring angular momentum back to the orbit. Due to the extreme mass ratio of the binary ($M_{\rm don}/M_{\rm acc} \ll 1$), the increase in the SMBH’s mass from the accreted material is negligible. 

In the extreme mass-ratio regime $M_{\rm don}/M_{\rm acc} \ll 1$, and for the assumptions about angular momentum transport considered in this work, $\dot{a}_{\rm MT}$ is negligible, and GW emission becomes the primary driver of orbital evolution, leading to an orbital decay that makes it harder for the mass transfer to remain stable.

In our calculations, we do not account for general relativistic corrections beyond the contribution from gravitational wave-driven orbital decay \citep{Peters_1963}. For most of the described evolution the star remains sufficiently distant from the SMBH that these corrections are negligible. Only during the final inspiral phase does the star approach the ISCO, where additional relativistic effects become significant. As a star approaches a distance of a few Schwarzschild radii, the MESA prescription may become insufficiently predictive for the final tens of years of the GW signal and the onset of last mass transfer phase. In particular, the relativistic tidal radius could be larger than its non-relativistic, Newtonian approximation, especially when accounting for partial disruption. The potential impact of the SMBH’s spin on the duration of the GW signal, as well as the star’s final fate, is discussed in Section \ref{section: gravitational-wave signal}.








\section{Results}

In Figure \ref{fig: Cartoon_stages}, we sketch four relevant evolutionary stages of the type of systems we simulate. The upper left panel is an illustration of the system formation via the Hills mechanism \citep{Hills1975}. A binary star system enters the Hill sphere of an SMBH and is disrupted. One star becomes bound to the SMBH on an eccentric orbit, while the other is ejected as a hypervelocity star. The upper right panel represents a stage when the bound star’s orbit shrinks and circularizes due to strong GW emission. The lower left panel illustrates a mass-transfer phase. As the orbit tightens, a subgiant fills its Roche lobe and undergoes stable mass transfer onto the SMBH, stripping the star of its hydrogen envelope. Finally, the lower right panel shows the GW-driven inspiral of the helium core. The stripped stellar core contracts and undergoes prolonged GW-driven orbital decay, eventually entering the LISA frequency band as a possibly detectable GW source.

\subsection{Mass and radius evolution}

\begin{figure} [ht]
    \centering
    \includegraphics[width=1.0\linewidth]{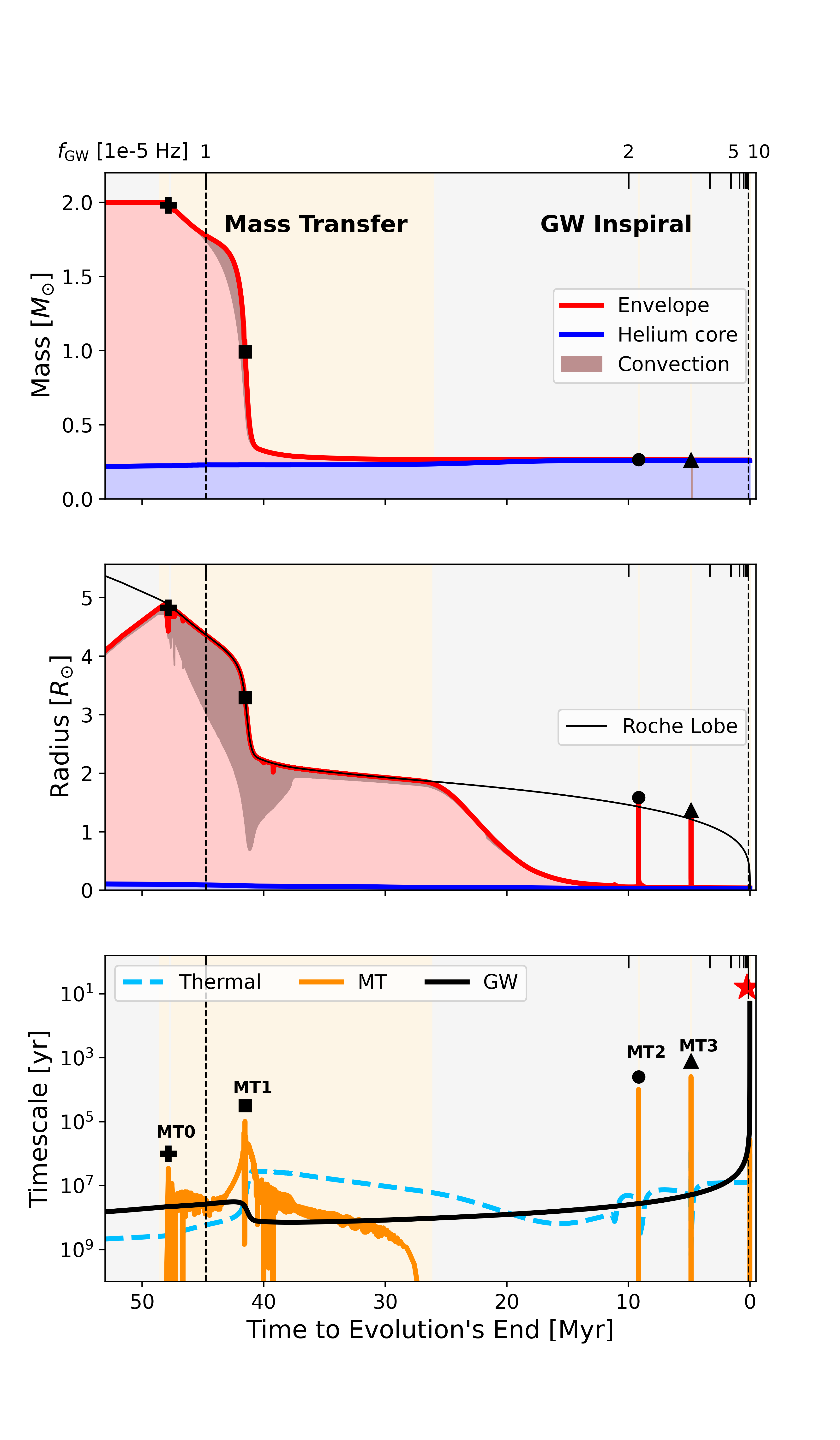}
    \caption{Evolution of a subgiant undergoing mass transfer as a function of time in the mass (upper panel) and radius (middle panel) coordinates. The internal structure of the subgiant is divided into regions: the hydrogen envelope (red) and the helium core (blue). The main convective regions are highlighted with shaded areas. The lower panel compares timescales for GW emission, $t_{\rm GW} = {a}/{\dot{a}_{\rm GW}}$ (gray), and mass transfer, $t_{\rm MT} =  {M_{\rm don}}/{\dot{M}}$ (orange). For comparison, we also plot the donor's thermal timescale (blue).  Different types of black markers have been used to highlight and localize characteristic features of mass transfer on three panels and in Figure \ref{fig: Lisa signal}. The final unstable mass transfer is marked by the red star.}
    \label{Fig: Mass_evolution}
\end{figure}

In Figure \ref{Fig: Mass_evolution}, we present results from {\tt MESA} for our radiative-envelope subgiant with \( M_{*} = 2 \, M_{\odot} \) and an \( M_{\bullet} = 4.3 \times 10^6 \, M_{\odot} \) SMBH companion. The figure includes the evolution of the mass (upper panel) and radius (middle panel) of the subgiant star during and after different mass-transfer phases. The lower panel shows the evolution of the timescales for GW emission, $t_{\rm GW} = {a}/{\dot{a}_{\rm GW}}$ (gray line), and mass transfer, $t_{\rm MT} =  {M_{\rm don}}/{\dot{M}}$ (orange line).\footnote{The mass-transfer timescale has been averaged over three-time steps to smooth the curve and reduce spikes of numerical origin.} We also plot the donor's thermal timescale (blue dashed line), approximated using $t_{\rm th} \approx 1.5 \times 10^7 (M_{\rm don}/M_{\odot})^2 (R_{\rm don}/R_{\odot})^{-1} (L_{\rm don}/L_{\odot})^{-1}\,\mathrm{yr}$ \citep{pols2011stellar}. Different types of black markers have been used to highlight and localize characteristic features of mass transfer across the three panels.

In the system we present, the subgiant initiates mass transfer once its helium core (which we define here as the region with hydrogen fraction X$<0.1$) has a mass $M_{\rm He} \approx 0.22 M_{\odot}$ (the first black marker on the left, labeled as `\textbf{MT0}'). First, the star undergoes a prolonged phase of stable mass transfer, lasting approximately $t_{\rm MT} \approx 20\,\rm Myr$. The mass transfer rate varies significantly over this time, ranging from $\dot{M}\sim 10^{-8}$ to $\sim 10^{-5} M_{\odot}$ yr$^{-1}$. During the first part of the mass transfer, the mass-transfer timescale is shorter than the estimated thermal timescale. Initially, a near-surface convective layer grows as the mass transfer proceeds and after $\sim 8 \, \rm Myr$, the rate achieves its peak at $\sim 10^{-5} M_{\odot}$ yr$^{-1}$ (the second black marker, labeled as `\textbf{MT1}'). This relatively short phase lasts $\sim 1.5\,\rm Myr$. After this rapid mass-transfer phase, half of the initial mass of the star ($\sim 1 M_{\odot}$) has been removed, and from now the depth of the outer convective layer shrinks in radial extent. Since then, the mass-transfer rate goes down systematically. This phase continues for another $\sim 13 \, \rm Myr$. The rate remains below $\dot{M}\sim 10^{-7} M_{\odot}$ yr$^{-1}$.

After the prolonged phase of stable mass transfer, nearly the entire convective hydrogen envelope of the star has been stripped, causing its radius to contract by a factor of $\sim 20$. The exposed helium core no longer fills its Roche lobe and enters a long-lasting ($\sim 15 \, \rm Myr$) GW-driven inspiral phase without further mass transfer. However, the spontaneous reignition of nuclear reactions in the remaining thin hydrogen shell, following the previous core contraction and heating of the outer layers, triggers a temporary but significant expansion of the star's radius. These thermal pulses lead to two additional short-lived mass-transfer phases (labeled as `\textbf{MT2}' and `\textbf{MT3}' in Fig. \ref{Fig: Mass_evolution}) with orbital periods of 20--30 hours, each lasting a few $10^4$ years. These phases are characterized by very high mass transfer rates of a few $\dot{M} \sim 10^{-5} M_{\odot}$ yr$^{-1}$. Notably, similar events have been observed involving a cooling proto-white dwarf, where reignition caused a temporary re-expansion into a cool giant phase \citep[most famously Sakurai's object, e.g.][in this case for helium reignition]{Duerbeck1996}. We further note that the evolution of this phase appears to proceed fast enough for the evolution to be directly observed (e.g., \citealt{Hajduk2005} for Sakurai's object, or \citealt{Istrate2016} for models in the regime of our example). In particular, we emphasize that flash-driven mass-transfer phases onto an SMBH may well switch on or off on human-observable timescales.


Following \cite{Dai2013}, we estimate the associated X-ray luminosity during the second and third mass-transfer phases(\textbf{MT2}, \textbf{MT3}) using the maximum Roche accretion power, given by $L_{\rm X} = \eta \dot{M} c^2$. Assuming a 10\% radiative efficiency for the accretion onto the SMBH, the $L_{\rm X}$ corresponding to the peak mass transfer rate of $\dot{M} \approx 6.8 \times 10^{-5} M_{\odot} \rm \, yr^{-1}$ is approximately $4 \times 10^{41} \rm erg \, s^{-1}$. We would like to emphasize that this work focuses on one example system. The thermal pulses, responsible for mass transfer and the resulting X-ray emission, could be associated with different orbital periods than the presented case (e.g., a few hours), producing multi-messenger signals in both the GW and EM spectra.

The orbital decay due to GW emission accelerates, eventually bringing the binary into frequencies above $f \gtrsim 5 \times 10^{-5} - 10^{-4} \, \rm{Hz}$, possibly entering the LISA band (depending on the distance of the source). From this moment, a system in the Galactic Center would emit GWs detectable by LISA (with SNR $\gtrapprox 7$) for hundreds of thousands of years. Ultimately, when the separation is very close to the SMBH's Schwarzschild radius, the compact helium core refills its Roche lobe. At this stage, we expect the mass transfer to become unstable. It develops a runaway process, which in the case of highly spinning SMBH may potentially lead to a partial disruption event in proximity to ISCO (see the discussion in Sec. \ref{section: gravitational-wave signal}). 

\subsection{Gravitational-wave signal in the LISA band and the final fate} \label{section: gravitational-wave signal}

In this section, we estimate the GW signal accompanying the evolution of the mass-transferring subgiant. The strain spectral density $h$ and approximations of the LISA sensitivity curve are calculated based on \cite{Robson2019}. For $h$, we use eq.~(27) of \cite{Robson2019}.



\begin{figure*}
    \hspace{1cm}
    \includegraphics[width=1.0\linewidth]{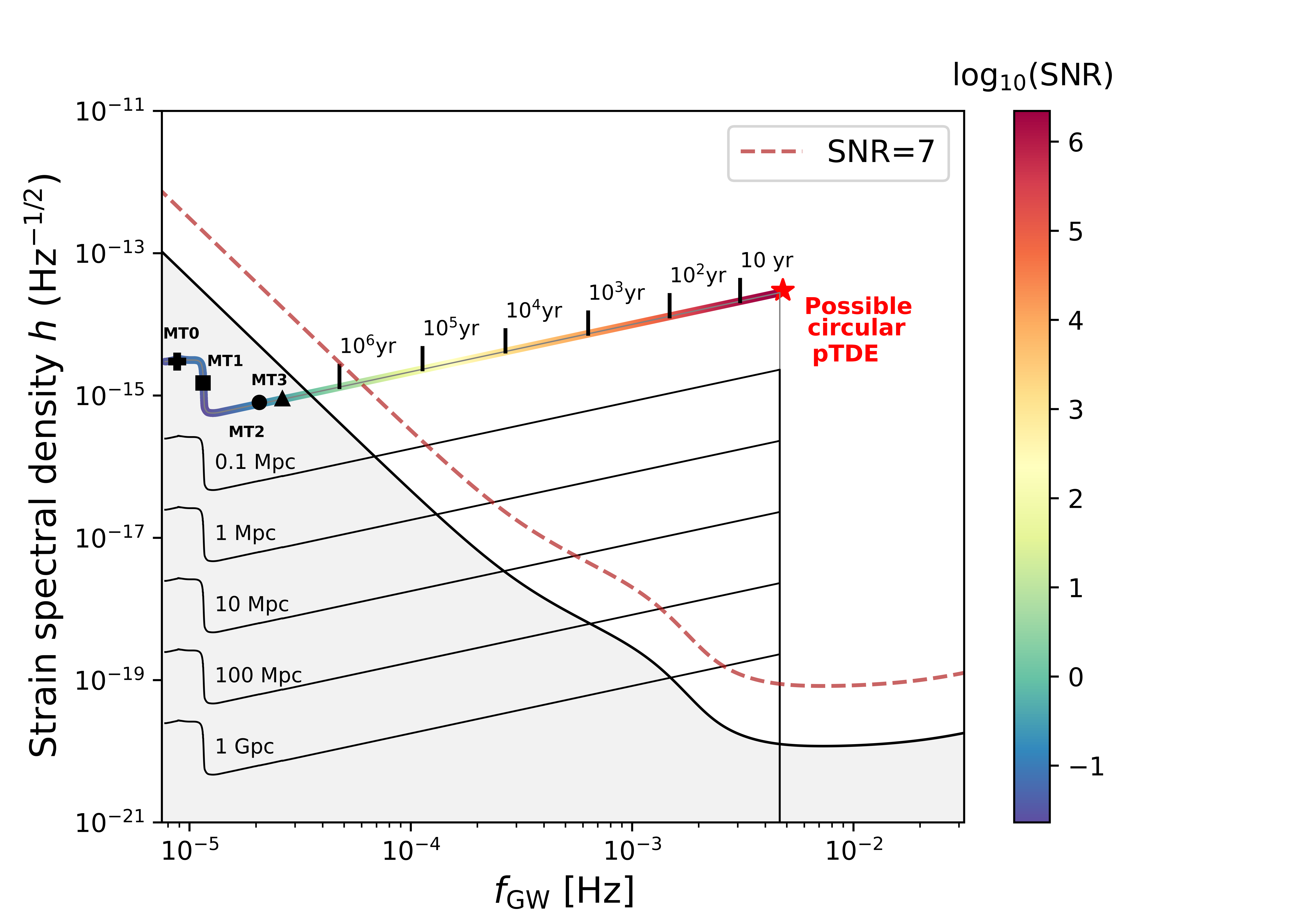}
    \caption{The strain spectral density $h$ of the GW signal from an initially $2  M_{\odot}$ subgiant transferring mass to a $4.3 \times 10^6 M_{\odot}$ SMBH, along with the LISA sensitivity curve for a 4-year mission \citep{Robson2019}. The colored line represents the signal if the system were located in the Milky Way, while the gray lines correspond to signals at greater distances. The numbers above the colored line indicate the time until the final unstable mass-transfer phase (red star) with the possible partial tidal disruption event (pTDE) on a circular orbit. The star undergoes its first mass-transfer phase at lower frequencies and subsequently enters the LISA band once it has been stripped of nearly its entire hydrogen envelope. Different types of black markers and labels: \textbf{MT0-MT4} have been used to localize and highlight characteristic features of mass transfer, consistently marked in Figure \ref{Fig: Mass_evolution}. At the final time step, the signal reaches a signal-to-noise ratio of $\sim 10^6$ in the Galactic Center, making it potentially detectable even from a distance of $1\,\rm Gpc$. }
    \label{fig: Lisa signal}
\end{figure*}

We plot the LISA sensitivity curve and the signal-to-noise ratio (SNR) following eq.~(1) of \citet{Robson2019}. The sensitivity curve $S_n(f_{\rm{GW}})$ components in Fig.~\ref{fig: Lisa signal} include the single-link optical metrology noise, the single test mass acceleration noise, and the Galactic confusion noise after 4 yr of the LISA mission (see Table 1 of \citealt{Robson2019}). These components are calculated using eqs.~(10), (11), and (15) from \citet{Robson2019}. The SNR for the basically monochromatic source is approximated as $\bar{\rho} \approx \frac{h^2(f_{\rm{GW}})}{S_n(f_{\rm{GW}})}$ following eq.~(26) of \cite{Robson2019}.

The GW signal obtained for our system in the frequency of the LISA band is shown in Fig.~\ref{fig: Lisa signal}. The colored line represents the signal if the system were located in the center of the Milky Way, while the gray lines correspond to signals at greater distances. The numbers above the colored line indicate the time until the final unstable mass-transfer phase and the potential partial tidal disruption. Characteristic mass transfer features, such as the start and the mass transfer peaks, are marked with black symbols in Figures \ref{Fig: Mass_evolution} and \ref{fig: Lisa signal} consistently. 

The subgiant undergoes its first long phase of mass transfer when the system is not yet detectable by LISA. The star loses most of its initial mass and nearly its entire hydrogen envelope. The reduction of mass by a factor of $\sim 10$ leads to a significant GW strain decrease after the mass-transfer phase. The star then undergoes two other shorter mass-transfer phases (lasting a few $10^4$ yr) caused by the significant radius expansion triggered by the spontaneous ignition of hydrogen layers. However, these phases, (indicated as \textbf{MT3} and \textbf{MT4}, and by black markers consistent with those in Figure \ref{Fig: Mass_evolution}) occur before the system enters the LISA frequency band. They do not significantly affect the mass of the star or the resulting GW signal. Subsequently, the star remains a compact helium core, and its radius contracts significantly (see Fig.~\ref{Fig: Mass_evolution}), allowing it to undergo a prolonged GW-driven inspiral. During this phase, the system enters the LISA band with a steadily increasing frequency $f{_{\rm GW}}$ and GW strain $(h \propto f^{2/3}_{\rm{GW}})$.  At the final time step, the signal reaches a signal-to-noise ratio of $\sim 10^6$ in the Galactic Center, making it potentially detectable even from a distance of $1\,\rm Gpc$. 

The GW-driven orbital decay rapidly accelerates (see the lower panel of Fig.~\ref{Fig: Mass_evolution}). At the final $\sim 10$ yr, the signal reaches an exceptionally high strain, making it potentially detectable even from a distance of $1\,\rm Gpc$.  

The Roche limit in the non-relativistic, Newtonian approximation for helium core, which is $a_{\rm Rl} \approx 19.8 R_{\odot}$ is very close to the Schwarzschild radius of the SMBH $R_{\rm schw} \approx 18.3 R_{\odot}$. Eventually, the separation between the SMBH and the helium core becomes so small that it refills its Roche lobe at $P\approx 8 \rm{min}$ and undergoes an unstable mass transfer phase.

We emphasize that the evolution of the system, particularly the final stages, is subject to significant uncertainty. At separations of $5-10 R_{\rm{schw}}$, the star's fate and therefore emitted GW signal depends on the SMBH's spin \citep{Dai2013}. Specifically, the size of the ISCO and, consequently, the duration of the GW-driven inspiral are influenced by the SMBH's rotation.  In the case of a prograde orbit, a higher SMBH spin leads to a smaller ISCO. As a result, the expected inspiral time is longer for a rapidly spinning SMBH compared to a slowly rotating one \citep[see, e.g., fig. 5 of][]{Dai2013}.

The observability of the final unstable mass transfer phase in the EM spectrum is determined by the spin of the SMBH and the structure of the helium core. Given the density of the core, only a partial disruption is expected rather than a complete one {Taeho2020}. A potential L2 outflow may occur if the ISCO lies within the tidal radius of the helium core. The partial tidal disruption radius would be larger than the Roche limit approximately by a factor of two {Taeho2020}, which may be further enhanced due to relativistic effects

The observability of the final unstable mass transfer phase in the EM spectrum is determined by the spin of the SMBH and the structure of the helium core. Given the density of the core, only a partial disruption is expected rather than a complete one \citep{Taeho2020}. A potential L2 outflow may occur if the ISCO lies within the tidal radius of the helium core. The partial tidal disruption radius would be larger than the Roche limit approximately by a factor of two \citep{Taeho2020}. It may be further enhanced due to relativistic effects \citep[see e.g.,][]{Dai2013} up to an another factor of two for the maximally spinning Kerr BHs. For a prograde orbit around a Kerr black hole, the ISCO is within $\sim 2 R_{\rm{schw}}$ (our tidal limit for a partial disruption) for a dimensionless SMBH spin parameter of $a \gtrsim 0.7$, see e.g., \cite{Bardeen1972}. Recent constraints on the spin of the Milky Way’s central SMBH suggest very high values of $a \approx 0.94$ \citep{EHTC2022}, making an observable disruption more likely.

Due to the star's proximity to the ISCO and its near-circular orbit, one can distinguish these events from classical TDEs by referring to them as extremely relativistic circular partial tidal disruptions \citep{Taeho2023}. The detection of such an event in the center of the Milky Way or another galaxy would suggest that the SMBH has a relatively high spin.

\subsection{Event rates and detection prospects}

In this section, we provide a rough estimate of the number of systems detectable by LISA in the local Universe. Our analysis focuses specifically on events involving stripped subgiants, similar to the example discussed earlier. While stellar EMRIs with main-sequence donors are expected to be more common due to their longer lifetimes and greater abundance, their evolutionary pathways and GW signals differ significantly. In particular, within the adopted mass transfer physics assumptions, the structure of main-sequence donors would prevent them from reaching the same high frequencies and SNR of GW as inspiraling stripped helium cores. Instead, the mass transfer would become unstable at wider separations. This topic will be addressed in a forthcoming study (Olejak et al., in prep.).

We calculate the expected number of systems per volume of detectable systems as follows:
\begin{equation}\label{eq. Nsys}
N_{\rm sys, V} \approx R_{\rm {form}} \times N_{\rm {gal, V}} \times t_{\rm LISA} \times  f_{\rm sub}
\end{equation}
where: $R_{\rm {form, V}}$ is the formation rate of subgiant star-SMBH EMRIs; $N_{\rm {gal}}$ is the number of galaxies per volume;
$t_{\rm dur}$ is the duration time a system is detectable by LISA with SNR$>7$; $R_{\rm {form}}$ is the rate of formation of stellar EMRIs with subgiant properties; $f_{\rm sub}$ is a value by which we need to reduce the rate considering only radiative-envelope subgiants.

\cite{LinialSari2023} estimates the formation rates of main sequence stellar EMRIs through two distinct channels:
via single-single scattering and the Hills mechanism. Both their prescriptions have been derived under the assumptions that the formation rate scales with the tidal radius of the star $r_{\text{tidal}}$ as $R_{\rm {form}} \sim 1/{r^2_{\text{tidal}}} \sim {1}/{r_{\text{star}}^2}$.  To adapt the prescription for subgiants, we must account for the fact that the radii of subgiants of our interest are 2~--~3 times larger than those of main sequence stars. Consequently, we need to reduce the rate for subgiant stars by introducing an additional multiplication factor of 0.2. Moreover, subgiants are significantly less common in the stellar population than main sequence stars. To estimate the fraction of subgiants with radiative envelopes in the stellar population, we must account for the distribution of stellar masses, lifetimes, and the duration of the subgiant phase. In particular, stars with masses that have lifetimes shorter than the age of the Universe make up approximately 10–15\% of the stellar population \citep{Salpeter1955}. On the other hand, the stars in proximity to SMBH do not necessarily follow the same mass distribution as stars in the field. Both theoretical studies \citep{Barak2025} and observations \citep{Mockler2022ApJ} indicate that the TDE rate for moderately massive stars ($M_{\star} \gtrsim 1.0 M_{\odot}$) is enhanced relative to their abundance in the overall stellar population. We assume that the mass range corresponding to stars with structures comparable to those of interest ($M_{\star}>1.2 m_{\odot}$), which could lead to similar evolutionary outcomes, constitutes approximately 10\% of the stellar population. The subgiant phase in which a star has not yet developed a deep convective envelope represents only a small fraction of a star’s lifetime. We assume this phase lasts at most 5\% of the total stellar lifetime. Consequently, we must scale our rates by another factor of $ \approx 0.5\,\%$. 

We estimate the formation rate using eqs.~\eqref{eq. R_scatt} and~\eqref{eq. R_Hills} of \cite{LinialSari2023} but modified for the case of subgiants by reducing values by a total factor of $f_{\rm sub} = 10^{-3}$.

We thereby obtain the formation rate via single-single scattering as:
\begin{align} \label{eq. R_scatt}
R_{\rm {scatt, sub}} = 10^{-3} \times  10^{-7} ({M_{\bullet}/{10^6 M_{\odot}}})^{1.1} M_{\star}^{-0.9} \text{yr}^{-1} \\
\approx 3 \times 10^{-10} \, \text{yr}^{-1}
\end{align}

\noindent and the formation rate via the Hills mechanism (disrupted binaries) assuming a binary fraction of $f_b = 50\,\%$ \citep{Offner2023}:
\begin{align} \label{eq. R_Hills}
R_{\rm {Hills, sub}} = 10^{-3} \times 10^{-5} \left( \frac{f_b}{0.1} \right) (M_{\bullet}/10^6 M_{\odot})^{-0.25} \, \text{yr}^{-1} \\
\approx 3 \times 10^{-8} \, \text{yr}^{-1}
\end{align}
where: $M_{\bullet}$ is the SMBH mass $M_\bullet=4.3 \times 10^6 M_{\odot}$; $f_b$ is the binary fraction; $M_{\star}$ is the mass of the star.

The total formation rate in Eq.~\eqref{eq. Nsys} is $R_{\rm form} = R_{\rm {scatt, sub}} + R_{\rm {Hills, sub}}$. The contribution from the Hills formation channel is significantly higher (by a factor of $\sim 100$), resulting in a formation rate of once per 300 million years in a Milky-Way-equivalent galaxy.

To estimate the number of systems, we need the number of local galaxies hosting an SMBH and the duration of the GW signal in the LISA band at various distances. Based on the results of large-scale galaxy surveys, we assume that there are $\sim 0.02$ galaxies hosting an SMBH per Mpc$^3$ \citep{Galaxy_Density2000,Graham2007}. This corresponds to around $2 \times 10^4$ galaxies within $100 \, \rm{Mpc}^3$ and $2 \times 10^7$ galaxies within $1 \, \rm{Gpc}^3$.

Based on the estimated GW signal in the LISA band (see Fig.~$\ref{fig: Lisa signal}$), we assume that the minimum duration for which a system is detectable within a volume of $1 \, \rm{Gpc}^3$ is 10 years. Therefore, the expected number of systems detectable with $\rm{SNR}>7$ within $1 \, \rm{Gpc}^3$ during the 4yr LISA mission is:
\begin{equation}
N_{\rm sys, 1Gpc} \approx (2 \times 10^7) \times 10\,{\rm yr} \times (3 \times 10^{-8}) \,{\rm yr^{-1}} = 6 .
\end{equation}

We can estimate the probability of hosting this kind of event in our own galactic center, expecting it will be detectable by LISA for $\sim 5 \times 10^5\, {\rm yr}$:
\begin{equation}
N_{\rm sys, MW} \approx 1 \times (5 \times 10^5\, {\rm yr}) \times (3 \times 10^{-8}) \,{\rm yr^{-1}} = 0.015 .
\end{equation}

The rates above are based on the theoretical predictions for the formation rates of stellar EMRIs. However, there are also hints of the existence of this kind of source among EM observations. In particular, new types of X-ray transients, QPEs, are expected to be caused by stars transferring mass on an SMBH \citep[e.g.,][]{LinialSari2023} or by the collision of a disk and a stellar-mass object \citep[e.g.,][]{Yao2024}. Recently, \citet{Arcodia2024} estimated the volumetric abundance for luminous QPEs as  $R_{\rm vol} =0.60^{+4.73}_{-0.43} \times 10^{-6}\,\rm Mpc^{-3}$.  Based on the volumetric rate of QPEs and the assumption of their stellar origin, \cite{Arcodia2024} also provides the lower limit for the expected number of stellar EMRI per year of $10^2-10^3$ within a volume of $z=1.0$. Under the assumption that comoving volume within $z=1$ is $V_c\approx100\,\rm Gpc^3$, the lower limit corresponds to 1~--~10 events per Gpc$^{3}$. This is in good agreement with our estimations, considering that the QPE events can also be caused by other objects, e.g., main sequence stars.

As regards detection prospects, we also note that the donor star in our example calculation has an orbital velocity of order $0.1c$ during the first three mass-transfer episodes.  So, in addition to other reasons for photometric variability due to the orbital motion, there may well be a substantial photometrically-detectable Doppler effect \citep[e.g.,][]{Shakura+Postnov1987, Zucker+2006}, as has already been observed for less extreme binaries \citep[e.g.,][]{Maxted+2000,vanKerkwijk+2010}, and discussed in the context of the S-stars by \citet{Rafikov2020}. In particular, during the two flash-driven mass-transfer phases, the model donor star has a luminosity of several hundred $L_{\odot}$ and a temperature around 20 kK (as shown in Fig.~\ref{Fig: HR}), with an orbital period of order a day. 

\section{Discussion} \label{sec: Discussion}

Our results are sensitive to the assumption regarding the amount of angular momentum transferred back to the orbit. Specifically, we obtain similar results for orbital evolution and the GW signal only if no more than $\sim 15\%$ of the angular momentum carried away by the mass transfer is returned to the orbit. If a larger fraction is transferred during mass transfer, the orbit with a subgiant donor widens significantly during rapid mass transfer instead of shrinking due to GW. This increase in separation between the star and the SMBH could lead to interactions with other stars or substantially extend the GW inspiral, preventing the system from entering the LISA detection band.

While the mechanism for transferring angular momentum back to the orbit remains unclear, some studies suggest that this process could be efficient, potentially enhancing the stability of mass transfer \citep[e.g.,][]{Dai2013, King2022, Wang2024}. Conversely, \citet{shu_structure_1979, pejcha_cool_2016,pejcha_binary_2016} have shown that for extreme mass ratios, the material lost from the outer Lagrange point L2 is marginally bound and can form either an inflated envelope around the binary or a circumbinary excretion disk. \citet{Linial2017} proposes that mass leakage through L2 would carry away extra energy from the orbit, causing the mass transfer to become more unstable. Our future work will explore alternative angular momentum loss scenarios and other evolutionary stages.

If an accretion disk forms from the star’s own material, it is likely to be coplanar with the stellar orbit. However, some disks could also originate from external material, such as debris from a previous TDE. In that case, the star may interact with this material, potentially affecting its structure. However, results from \citet{Yao2024} for main sequence star donors suggest that such collisions primarily heat the outer layers of the star without significantly disrupting its inner structure. For subgiants, we expect the impact on the stellar core to be even less pronounced due to their more centrally concentrated structure.

It is important to note that this study focuses on scenarios in quiescent galaxies, where stellar EMRIs form via the Hills mechanism or single-single scattering. The formation and evolution of stellar EMRIs in AGN disks are subjects for future work. In particular, one would need to account for specific boundary conditions, as stars in AGN disks exist in a much hotter and denser environment \citep[e.g.][]{Cantiello2021}.

Another consideration is that the orbit may remain mildly eccentric at the onset of mass transfer. Eccentricity could lead to tidal heating, injecting additional energy into the star’s outer layers and causing it to expand \citep[e.g.,][]{Syer+1991, Quataert-KITP-2024}. This expansion might trigger the onset of mass transfer at a greater distance from the SMBH, potentially increasing stability under our assumptions. However, due to the subgiant's structure, this would not significantly affect the primary result—the GW signal detectable by LISA—since the star would enter the detection band only after the mass-transfer phase. 

For extragalactic sources, it is important to account for the fact that the size of the ISCO varies with the mass of the SMBH. This variation influences the maximum frequencies and strain of the GW signal emitted by the system during its final stages. The location of the ISCO and, consequently, the final phase of the GW inspiral are determined by the spin of the SMBH \citep{Dai2013}.

\section{Conclusions}

We present a model of a subgiant star with radiative envelope transferring mass to a supermassive SMBH using the {\tt MESA} stellar evolution code. Our study focuses on the stability of the mass transfer phase, presenting a case of $2 M_{\odot}$ subgiant ultimately stripped down to a hot helium core. We characterize the resulting GW signal and the potential detectability of such systems as a bright multi-messenger source for the future LISA mission. We find that:

\begin{itemize}
    \item Subgiants undergo their first mass transfer before entering the LISA bandwidth. Mass transfer from the subgiant star to an SMBH remains stable for approximately $\sim 10^6$ years. This stability holds even if no angular momentum from the transferred material is returned to the orbit to counterbalance the GW-driven orbital decay.
    \item These systems can become exceptionally bright and long-lasting GW sources for LISA, after the stripped subgiant’s radius contracts. The remaining helium core goes through a prolonged GW inspiral, with both the signal frequency and strain increasing over time. Such a system would be a bright source for LISA, possibly detectable from large distances up to 1 Gpc covering, e.g., the Abell clusters.
    \item We estimate that a few systems involving similar stripped subgiant stars could be detectable by LISA within a volume of up to 1 Gpc$^3$ during a 4\,yr mission duration.
    \item There is a 1$\%$ chance of hosting such an event in our galactic center, spending several $\sim 10^5$yr in the bandwidth of LISA. A GW signal from a system in the Milky Way could achieve extreme signal-to-noise ratios of around $10^6$.
    \item As the helium core evolves, hydrogen flashes in the thin remaining envelope layers trigger a few additional rapid mass-transfer phases at orbital periods of $\sim 20-30$ hours. We estimate their associated X-ray luminosity $L_{\rm X}$, assuming a 10\% radiative efficiency for the accretion onto the SMBH, to be a few $\sim 10^{41} \rm erg \, s^{-1}$. Eventually, the separation decreases below the Roche limit of the helium core, which is slightly bigger than the Schwarzschild radius of the SMBH. In the case of high SMBH spin (a$>0.7$), this final, unstable mass-transfer phase could lead to partial tidal disruption on a circular orbit associated with the orbital period of $\sim 10$ minutes. 
    Further work is needed to determine whether and what kinds of observed EM transients might be generated in galactic centers by described types of mass transfer phases. The studied systems are potential multi-messenger sources, detectable through both EM and GW emission—either simultaneously or at different stages of their evolution.
\end{itemize}

This study highlights the potential of stellar EMRIs as multi-messenger sources, with a particular emphasis on their viability as bright LISA detections. Future detection of such a system in GW or the EM spectrum would provide valuable insights into stellar astrophysics, the physics of mass transfer, and the dynamics of galactic centers.



\section{Acknowledgments}
The authors thank Taeho Ryu, Riccardo Arcodia, Norbert Langer, Marco Antonelli, Evgeni Grishin, Abinaya Swaruba Rajamuthukumar, and Lazaros Souvaitzis for their valuable comments. SdM acknowledges discussions with Elliot Quataert and Philippe Yao. AO, JS, SDM, RV, and SJ acknowledge funding from the Netherlands Organization for Scientific Research (NWO), as part of the Vidi research program BinWaves (project number 639.042.728, PI: de Mink). This research was supported in part by grant NSF PHY-2309135 to the Kavli Institute for Theoretical Physics (KITP).

%

\vspace{5mm}

\software{Python matplotlib, {\tt MESA}}



\appendix
\section{Evolution on HR diagram}

\begin{figure} 
    \centering
    \includegraphics[width=0.55\linewidth]{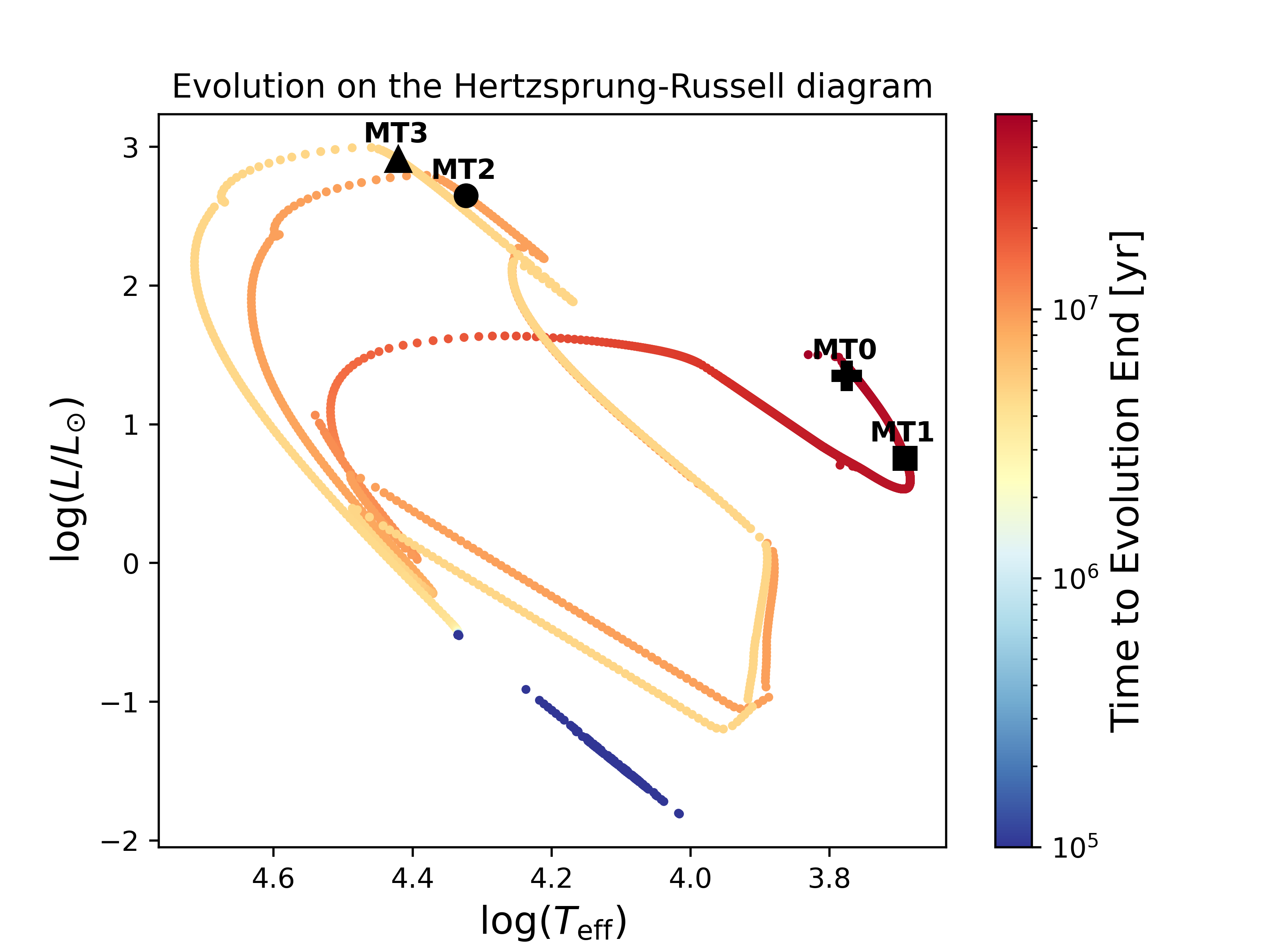}
    \caption{Evolution of the system on Hertzsprung-Russell (HR) diagram. Different types of black markers and labels: MT0-MT4 have been used to localize and highlight characteristic features of mass transfer, consistently marked in Figure \ref{Fig: Mass_evolution} and \ref{fig: Lisa signal}.} 
    \label{Fig: HR}
\end{figure} 

\bibliography{sample631}{}

\begin{thebibliography}{}
\expandafter\ifx\csname natexlab\endcsname\relax\def\natexlab#1{#1}\fi
\providecommand{\url}[1]{\href{#1}{#1}}
\providecommand{\dodoi}[1]{doi:~\href{http://doi.org/#1}{\nolinkurl{#1}}}
\providecommand{\doeprint}[1]{\href{http://ascl.net/#1}{\nolinkurl{http://ascl.net/#1}}}
\providecommand{\doarXiv}[1]{\href{https://arxiv.org/abs/#1}{\nolinkurl{https://arxiv.org/abs/#1}}}

\bibitem[{{Amaro-Seoane} {et~al.}(2007){Amaro-Seoane}, {Gair}, {Freitag},
  {Miller}, {Mandel}, {Cutler}, \& {Babak}}]{Amaro-Seoane2007}
{Amaro-Seoane}, P., {Gair}, J.~R., {Freitag}, M., {et~al.} 2007, Classical and
  Quantum Gravity, 24, R113, \dodoi{10.1088/0264-9381/24/17/R01}

\bibitem[{{Amaro-Seoane} {et~al.}(2017){Amaro-Seoane}, {Audley}, {Babak},
  {Baker}, {Barausse}, {Bender}, {Berti}, {Binetruy}, {Born}, {Bortoluzzi},
  {Camp}, {Caprini}, {Cardoso}, {Colpi}, {Conklin}, {Cornish}, {Cutler},
  {Danzmann}, {Dolesi}, {Ferraioli}, {Ferroni}, {Fitzsimons}, {Gair}, {Gesa
  Bote}, {Giardini}, {Gibert}, {Grimani}, {Halloin}, {Heinzel}, {Hertog},
  {Hewitson}, {Holley-Bockelmann}, {Hollington}, {Hueller}, {Inchauspe},
  {Jetzer}, {Karnesis}, {Killow}, {Klein}, {Klipstein}, {Korsakova}, {Larson},
  {Livas}, {Lloro}, {Man}, {Mance}, {Martino}, {Mateos}, {McKenzie},
  {McWilliams}, {Miller}, {Mueller}, {Nardini}, {Nelemans}, {Nofrarias},
  {Petiteau}, {Pivato}, {Plagnol}, {Porter}, {Reiche}, {Robertson},
  {Robertson}, {Rossi}, {Russano}, {Schutz}, {Sesana}, {Shoemaker}, {Slutsky},
  {Sopuerta}, {Sumner}, {Tamanini}, {Thorpe}, {Troebs}, {Vallisneri},
  {Vecchio}, {Vetrugno}, {Vitale}, {Volonteri}, {Wanner}, {Ward}, {Wass},
  {Weber}, {Ziemer}, \& {Zweifel}}]{AmaroSeoane2017}
{Amaro-Seoane}, P., {Audley}, H., {Babak}, S., {et~al.} 2017, arXiv e-prints,
  arXiv:1702.00786, \dodoi{10.48550/arXiv.1702.00786}

\bibitem[{{Amaro-Seoane} {et~al.}(2023){Amaro-Seoane}, {Andrews}, {Arca Sedda},
  {Askar}, {Baghi}, {Balasov}, {Bartos}, {Bavera}, {Bellovary}, {Berry},
  {Berti}, {Bianchi}, {Blecha}, {Blondin}, {Bogdanovi{\'c}}, {Boissier},
  {Bonetti}, {Bonoli}, {Bortolas}, {Breivik}, {Capelo}, {Caramete},
  {Cattorini}, {Charisi}, {Chaty}, {Chen}, {Chru{\'s}li{\'n}ska}, {Chua},
  {Church}, {Colpi}, {D'Orazio}, {Danielski}, {Davies}, {Dayal}, {De Rosa},
  {Derdzinski}, {Destounis}, {Dotti}, {Du{\c{t}}an}, {Dvorkin}, {Fabj},
  {Foglizzo}, {Ford}, {Fouvry}, {Franchini}, {Fragos}, {Fryer}, {Gaspari},
  {Gerosa}, {Graziani}, {Groot}, {Habouzit}, {Haggard}, {Haiman}, {Han},
  {Istrate}, {Johansson}, {Khan}, {Kimpson}, {Kokkotas}, {Kong}, {Korol},
  {Kremer}, {Kupfer}, {Lamberts}, {Larson}, {Lau}, {Liu}, {Lloyd-Ronning},
  {Lodato}, {Lupi}, {Ma}, {Maccarone}, {Mandel}, {Mangiagli}, {Mapelli},
  {Mathis}, {Mayer}, {McGee}, {McKernan}, {Miller}, {Mota}, {Mumpower},
  {Nasim}, {Nelemans}, {Noble}, {Pacucci}, {Panessa}, {Paschalidis}, {Pfister},
  {Porquet}, {Quenby}, {Ricarte}, {R{\"o}pke}, {Regan}, {Rosswog}, {Ruiter},
  {Ruiz}, {Runnoe}, {Schneider}, {Schnittman}, {Secunda}, {Sesana}, {Seto},
  {Shao}, {Shapiro}, {Sopuerta}, {Stone}, {Suvorov}, {Tamanini}, {Tamfal},
  {Tauris}, {Temmink}, {Tomsick}, {Toonen}, {Torres-Orjuela}, {Toscani},
  {Tsokaros}, {Unal}, {V{\'a}zquez-Aceves}, {Valiante}, {van Putten}, {van
  Roestel}, {Vignali}, {Volonteri}, {Wu}, {Younsi}, {Yu}, {Zane}, {Zwick},
  {Antonini}, {Baibhav}, {Barausse}, {Bonilla Rivera}, {Branchesi},
  {Branduardi-Raymont}, {Burdge}, {Chakraborty}, {Cuadra}, {Dage}, {Davis}, {de
  Mink}, {Decarli}, {Doneva}, {Escoffier}, {Gandhi}, {Haardt}, {Lousto},
  {Nissanke}, {Nordhaus}, {O'Shaughnessy}, {Portegies Zwart}, {Pound},
  {Schussler}, {Sergijenko}, {Spallicci}, {Vernieri}, \&
  {Vigna-G{\'o}mez}}]{LISA2023}
{Amaro-Seoane}, P., {Andrews}, J., {Arca Sedda}, M., {et~al.} 2023, Living
  Reviews in Relativity, 26, 2, \dodoi{10.1007/s41114-022-00041-y}

\bibitem[{{Arcodia} {et~al.}(2021){Arcodia}, {Merloni}, {Nandra}, {Buchner},
  {Salvato}, {Pasham}, {Remillard}, {Comparat}, {Lamer}, {Ponti}, {Malyali},
  {Wolf}, {Arzoumanian}, {Bogensberger}, {Buckley}, {Gendreau}, {Gromadzki},
  {Kara}, {Krumpe}, {Markwardt}, {Ramos-Ceja}, {Rau}, {Schramm}, \&
  {Schwope}}]{Arcodia2021}
{Arcodia}, R., {Merloni}, A., {Nandra}, K., {et~al.} 2021, \nat, 592, 704,
  \dodoi{10.1038/s41586-021-03394-6}

\bibitem[{{Arcodia} {et~al.}(2024){Arcodia}, {Linial}, {Miniutti}, {Franchini},
  {Giustini}, {Bonetti}, {Sesana}, {Soria}, {Chakraborty}, {Dotti}, {Kara},
  {Merloni}, {Ponti}, \& {Vincentelli}}]{Arcodia2024}
{Arcodia}, R., {Linial}, I., {Miniutti}, G., {et~al.} 2024, arXiv e-prints,
  arXiv:2406.17020, \dodoi{10.48550/arXiv.2406.17020}

\bibitem[{{Bardeen} {et~al.}(1972){Bardeen}, {Press}, \&
  {Teukolsky}}]{Bardeen1972}
{Bardeen}, J.~M., {Press}, W.~H., \& {Teukolsky}, S.~A. 1972, \apj, 178, 347,
  \dodoi{10.1086/151796}

\bibitem[{{Bykov} {et~al.}(2024){Bykov}, {Gilfanov}, {Sunyaev}, \&
  {Medvedev}}]{Bykov2024}
{Bykov}, S., {Gilfanov}, M., {Sunyaev}, R., \& {Medvedev}, P. 2024, arXiv
  e-prints, arXiv:2409.16908, \dodoi{10.48550/arXiv.2409.16908}

\bibitem[{{Cantiello} {et~al.}(2021){Cantiello}, {Jermyn}, \&
  {Lin}}]{Cantiello2021}
{Cantiello}, M., {Jermyn}, A.~S., \& {Lin}, D. N.~C. 2021, \apj, 910, 94,
  \dodoi{10.3847/1538-4357/abdf4f}

\bibitem[{{Chakraborty} {et~al.}(2021){Chakraborty}, {Kara}, {Masterson},
  {Giustini}, {Miniutti}, \& {Saxton}}]{Chakraborty2021}
{Chakraborty}, J., {Kara}, E., {Masterson}, M., {et~al.} 2021, \apjl, 921, L40,
  \dodoi{10.3847/2041-8213/ac313b}

\bibitem[{{Chen} {et~al.}(2022){Chen}, {Qiu}, {Li}, \& {Liu}}]{Chen2022}
{Chen}, X., {Qiu}, Y., {Li}, S., \& {Liu}, F.~K. 2022, \apj, 930, 122,
  \dodoi{10.3847/1538-4357/ac63bf}

\bibitem[{{Colpi} {et~al.}(2019){Colpi}, {Holley-Bockelmann}, {Bogdanovic},
  {Natarajan}, {Bellovary}, {Sesana}, {Tremmel}, {Schnittman}, {Comerford},
  {Barausse}, {Berti}, {Volonteri}, {Khan}, {McWilliams}, {Burke-Spolaor},
  {Hazboun}, {Conklin}, {Mueller}, \& {Larson}}]{Colpi2019}
{Colpi}, M., {Holley-Bockelmann}, K., {Bogdanovic}, T., {et~al.} 2019, arXiv
  e-prints, arXiv:1903.06867, \dodoi{10.48550/arXiv.1903.06867}

\bibitem[{{Dai} \& {Blandford}(2013)}]{Dai2013}
{Dai}, L., \& {Blandford}, R. 2013, \mnras, 434, 2948,
  \dodoi{10.1093/mnras/stt1209}

\bibitem[{{D'Orazio} {et~al.}(2025){D'Orazio}, {Tiede}, {Zwick}, {Hayasaki}, \&
  {Mayer}}]{dorazio2025}
{D'Orazio}, D.~J., {Tiede}, C., {Zwick}, L., {Hayasaki}, K., \& {Mayer}, L.
  2025, arXiv e-prints, arXiv:2501.10509.
\newblock \doarXiv{2501.10509}

\bibitem[{{Duerbeck} \& {Benetti}(1996)}]{Duerbeck1996}
{Duerbeck}, H.~W., \& {Benetti}, S. 1996, \apjl, 468, L111,
  \dodoi{10.1086/310241}

\bibitem[{{Eggleton}(1983)}]{Eggleton1983}
{Eggleton}, P.~P. 1983, \apj, 268, 368, \dodoi{10.1086/160960}

\bibitem[{{Event Horizon Telescope Collaboration} {et~al.}(2022){Event Horizon
  Telescope Collaboration}, {Akiyama}, {Alberdi}, {Alef}, {Algaba}, {Anantua},
  {Asada}, {Azulay}, {Bach}, {Baczko}, {Ball}, {Balokovi{\'c}}, {Barrett},
  {Baub{\"o}ck}, {Benson}, {Bintley}, {Blackburn}, {Blundell}, {Bouman},
  {Bower}, {Boyce}, {Bremer}, {Brinkerink}, {Brissenden}, {Britzen},
  {Broderick}, {Broguiere}, {Bronzwaer}, {Bustamante}, {Byun}, {Carlstrom},
  {Ceccobello}, {Chael}, {Chan}, {Chatterjee}, {Chatterjee}, {Chen}, {Chen},
  {Cheng}, {Cho}, {Christian}, {Conroy}, {Conway}, {Cordes}, {Crawford},
  {Crew}, {Cruz-Osorio}, {Cui}, {Davelaar}, {De Laurentis}, {Deane}, {Dempsey},
  {Desvignes}, {Dexter}, {Dhruv}, {Doeleman}, {Dougal}, {Dzib}, {Eatough},
  {Emami}, {Falcke}, {Farah}, {Fish}, {Fomalont}, {Ford}, {Fraga-Encinas},
  {Freeman}, {Friberg}, {Fromm}, {Fuentes}, {Galison}, {Gammie}, {Garc{\'\i}a},
  {Gentaz}, {Georgiev}, {Goddi}, {Gold}, {G{\'o}mez-Ruiz}, {G{\'o}mez}, {Gu},
  {Gurwell}, {Hada}, {Haggard}, {Haworth}, {Hecht}, {Hesper}, {Heumann}, {Ho},
  {Ho}, {Honma}, {Huang}, {Huang}, {Hughes}, {Ikeda}, {Impellizzeri}, {Inoue},
  {Issaoun}, {James}, {Jannuzi}, {Janssen}, {Jeter}, {Jiang},
  {Jim{\'e}nez-Rosales}, {Johnson}, {Jorstad}, {Joshi}, {Jung}, {Karami},
  {Karuppusamy}, {Kawashima}, {Keating}, {Kettenis}, {Kim}, {Kim}, {Kim},
  {Kim}, {Kino}, {Koay}, {Kocherlakota}, {Kofuji}, {Koch}, {Koyama}, {Kramer},
  {Kramer}, {Krichbaum}, {Kuo}, {La Bella}, {Lauer}, {Lee}, {Lee}, {Leung},
  {Levis}, {Li}, {Lico}, {Lindahl}, {Lindqvist}, {Lisakov}, {Liu}, {Liu},
  {Liuzzo}, {Lo}, {Lobanov}, {Loinard}, {Lonsdale}, {Lu}, {Mao}, {Marchili},
  {Markoff}, {Marrone}, {Marscher}, {Mart{\'\i}-Vidal}, {Matsushita},
  {Matthews}, {Medeiros}, {Menten}, {Michalik}, {Mizuno}, {Mizuno}, {Moran},
  {Moriyama}, {Moscibrodzka}, {M{\"u}ller}, {Mus}, {Musoke}, {Myserlis},
  {Nadolski}, {Nagai}, {Nagar}, {Nakamura}, {Narayan}, {Narayanan},
  {Natarajan}, {Nathanail}, {Fuentes}, {Neilsen}, {Neri}, {Ni}, {Noutsos},
  {Nowak}, {Oh}, {Okino}, {Olivares}, {Ortiz-Le{\'o}n}, {Oyama}, {{\"O}zel},
  {Palumbo}, {Paraschos}, {Park}, {Parsons}, {Patel}, {Pen}, {Pesce},
  {Pi{\'e}tu}, {Plambeck}, {PopStefanija}, {Porth}, {P{\"o}tzl}, {Prather},
  {Preciado-L{\'o}pez}, \& {Psaltis}}]{EHTC2022}
{Event Horizon Telescope Collaboration}, {Akiyama}, K., {Alberdi}, A., {et~al.}
  2022, \apjl, 930, L12, \dodoi{10.3847/2041-8213/ac6674}

\bibitem[{{Franchini} {et~al.}(2023){Franchini}, {Bonetti}, {Lupi}, {Miniutti},
  {Bortolas}, {Giustini}, {Dotti}, {Sesana}, {Arcodia}, \&
  {Ryu}}]{Franchini2023}
{Franchini}, A., {Bonetti}, M., {Lupi}, A., {et~al.} 2023, \aap, 675, A100,
  \dodoi{10.1051/0004-6361/202346565}

\bibitem[{{Gezari}(2021)}]{Gezari2021}
{Gezari}, S. 2021, \araa, 59, 21, \dodoi{10.1146/annurev-astro-111720-030029}

\bibitem[{{Ghez} {et~al.}(2003){Ghez}, {Duch{\^e}ne}, {Matthews}, {Hornstein},
  {Tanner}, {Larkin}, {Morris}, {Becklin}, {Salim}, {Kremenek}, {Thompson},
  {Soifer}, {Neugebauer}, \& {McLean}}]{Ghez2003}
{Ghez}, A.~M., {Duch{\^e}ne}, G., {Matthews}, K., {et~al.} 2003, \apjl, 586,
  L127, \dodoi{10.1086/374804}

\bibitem[{{Giustini} {et~al.}(2020){Giustini}, {Miniutti}, \&
  {Saxton}}]{Giustini2020}
{Giustini}, M., {Miniutti}, G., \& {Saxton}, R.~D. 2020, \aap, 636, L2,
  \dodoi{10.1051/0004-6361/202037610}

\bibitem[{{Gong} {et~al.}(2021){Gong}, {Luo}, \& {Wang}}]{ChineseGW2021}
{Gong}, Y., {Luo}, J., \& {Wang}, B. 2021, Nature Astronomy, 5, 881,
  \dodoi{10.1038/s41550-021-01480-3}

\bibitem[{{Graham} {et~al.}(2007){Graham}, {Driver}, {Allen}, \&
  {Liske}}]{Graham2007}
{Graham}, A.~W., {Driver}, S.~P., {Allen}, P.~D., \& {Liske}, J. 2007, \mnras,
  378, 198, \dodoi{10.1111/j.1365-2966.2007.11770.x}

\bibitem[{{Gravity Collaboration} {et~al.}(2023){Gravity Collaboration},
  {Abuter}, {Aimar}, {Amaro Seoane}, {Amorim}, {Baub{\"o}ck}, {Berger},
  {Bonnet}, {Bourdarot}, {Brandner}, {Cardoso}, {Cl{\'e}net}, {Davies}, {de
  Zeeuw}, {Dexter}, {Drescher}, {Eckart}, {Eisenhauer}, {Feuchtgruber},
  {Finger}, {F{\"o}rster Schreiber}, {Foschi}, {Garcia}, {Gao}, {Gelles},
  {Gendron}, {Genzel}, {Gillessen}, {Hartl}, {Haubois}, {Haussmann},
  {Hei{\ss}el}, {Henning}, {Hippler}, {Horrobin}, {Jochum}, {Jocou}, {Kaufer},
  {Kervella}, {Lacour}, {Lapeyr{\`e}re}, {Le Bouquin}, {L{\'e}na}, {Lutz},
  {Mang}, {More}, {Ott}, {Paumard}, {Perraut}, {Perrin}, {Pfuhl}, {Rabien},
  {Ribeiro}, {Sadun Bordoni}, {Scheithauer}, {Shangguan}, {Shimizu}, {Stadler},
  {Straub}, {Straubmeier}, {Sturm}, {Tacconi}, {Vincent}, {von Fellenberg},
  {Widmann}, {Wielgus}, {Wieprecht}, {Wiezorrek}, \& {Woillez}}]{Gravity2023}
{Gravity Collaboration}, {Abuter}, R., {Aimar}, N., {et~al.} 2023, \aap, 677,
  L10, \dodoi{10.1051/0004-6361/202347416}

\bibitem[{{Hajduk} {et~al.}(2005){Hajduk}, {Zijlstra}, {Herwig}, {van Hoof},
  {Kerber}, {Kimeswenger}, {Pollacco}, {Evans}, {Lop{\'e}z}, {Bryce}, {Eyres},
  \& {Matsuura}}]{Hajduk2005}
{Hajduk}, M., {Zijlstra}, A.~A., {Herwig}, F., {et~al.} 2005, Science, 308,
  231, \dodoi{10.1126/science.1108953}

\bibitem[{{Hameury} {et~al.}(1994){Hameury}, {King}, {Lasota}, \&
  {Auvergne}}]{Hameury1994}
{Hameury}, J.~M., {King}, A.~R., {Lasota}, J.~P., \& {Auvergne}, M. 1994, \aap,
  292, 404

\bibitem[{{Hills}(1975)}]{Hills1975}
{Hills}, J.~G. 1975, \nat, 254, 295, \dodoi{10.1038/254295a0}

\bibitem[{{Istrate} {et~al.}(2016){Istrate}, {Marchant}, {Tauris}, {Langer},
  {Stancliffe}, \& {Grassitelli}}]{Istrate2016}
{Istrate}, A.~G., {Marchant}, P., {Tauris}, T.~M., {et~al.} 2016, \aap, 595,
  A35, \dodoi{10.1051/0004-6361/201628874}

\bibitem[{{Jermyn} {et~al.}(2023){Jermyn}, {Bauer}, {Schwab}, {Farmer}, {Ball},
  {Bellinger}, {Dotter}, {Joyce}, {Marchant}, {Mombarg}, {Wolf}, {Sunny Wong},
  {Cinquegrana}, {Farrell}, {Smolec}, {Thoul}, {Cantiello}, {Herwig}, {Toloza},
  {Bildsten}, {Townsend}, \& {Timmes}}]{Jermyn2023}
{Jermyn}, A.~S., {Bauer}, E.~B., {Schwab}, J., {et~al.} 2023, \apjs, 265, 15,
  \dodoi{10.3847/1538-4365/acae8d}

\bibitem[{{Kejriwal} {et~al.}(2024){Kejriwal}, {Witzany}, {Zaja{\v{c}}ek},
  {Pasham}, \& {Chua}}]{Kejriwal2024}
{Kejriwal}, S., {Witzany}, V., {Zaja{\v{c}}ek}, M., {Pasham}, D.~R., \& {Chua},
  A. J.~K. 2024, \mnras, 532, 2143, \dodoi{10.1093/mnras/stae1599}

\bibitem[{{King}(2020)}]{King2020}
{King}, A. 2020, \mnras, 493, L120, \dodoi{10.1093/mnrasl/slaa020}

\bibitem[{{King}(2022)}]{King2022}
---. 2022, \mnras, 515, 4344, \dodoi{10.1093/mnras/stac1641}

\bibitem[{{Kolb} \& {Ritter}(1990)}]{Kolb1990}
{Kolb}, U., \& {Ritter}, H. 1990, \aap, 236, 385

\bibitem[{{Krolik} \& {Linial}(2022)}]{Krolik2022}
{Krolik}, J.~H., \& {Linial}, I. 2022, \apj, 941, 24,
  \dodoi{10.3847/1538-4357/ac9eb6}

\bibitem[{{Linial} \& {Metzger}(2023)}]{Linial2023}
{Linial}, I., \& {Metzger}, B.~D. 2023, \apj, 957, 34,
  \dodoi{10.3847/1538-4357/acf65b}

\bibitem[{{Linial} \& {Sari}(2017)}]{Linial2017}
{Linial}, I., \& {Sari}, R. 2017, \mnras, 469, 2441,
  \dodoi{10.1093/mnras/stx1041}

\bibitem[{{Linial} \& {Sari}(2023)}]{LinialSari2023}
---. 2023, \apj, 945, 86, \dodoi{10.3847/1538-4357/acbd3d}

\bibitem[{{Lu} \& {Quataert}(2023)}]{Lu2023}
{Lu}, W., \& {Quataert}, E. 2023, \mnras, 524, 6247,
  \dodoi{10.1093/mnras/stad2203}

\bibitem[{{Luo} {et~al.}(2016){Luo}, {Chen}, {Duan}, {Gong}, {Hu}, {Ji}, {Liu},
  {Mei}, {Milyukov}, {Sazhin}, {Shao}, {Toth}, {Tu}, {Wang}, {Wang}, {Yeh},
  {Zhan}, {Zhang}, {Zharov}, \& {Zhou}}]{Luo2016}
{Luo}, J., {Chen}, L.-S., {Duan}, H.-Z., {et~al.} 2016, Classical and Quantum
  Gravity, 33, 035010, \dodoi{10.1088/0264-9381/33/3/035010}

\bibitem[{{Lyu} {et~al.}(2024){Lyu}, {Pan}, {Mao}, {Jiang}, \&
  {Yang}}]{Lyu2025}
{Lyu}, Z., {Pan}, Z., {Mao}, J., {Jiang}, N., \& {Yang}, H. 2024, arXiv
  e-prints, arXiv:2501.03252.
\newblock \doarXiv{2501.03252}

\bibitem[{{Magorrian} \& {Tremaine}(1999)}]{Magorrian1999}
{Magorrian}, J., \& {Tremaine}, S. 1999, \mnras, 309, 447,
  \dodoi{10.1046/j.1365-8711.1999.02853.x}

\bibitem[{{Maxted} {et~al.}(2000){Maxted}, {Marsh}, \& {North}}]{Maxted+2000}
{Maxted}, P.~F.~L., {Marsh}, T.~R., \& {North}, R.~C. 2000, \mnras, 317, L41,
  \dodoi{10.1046/j.1365-8711.2000.03856.x}

\bibitem[{{Metzger} {et~al.}(2022){Metzger}, {Stone}, \&
  {Gilbaum}}]{Metzger2022}
{Metzger}, B.~D., {Stone}, N.~C., \& {Gilbaum}, S. 2022, \apj, 926, 101,
  \dodoi{10.3847/1538-4357/ac3ee1}

\bibitem[{{Miniutti} {et~al.}(2019){Miniutti}, {Saxton}, {Giustini},
  {Alexander}, {Fender}, {Heywood}, {Monageng}, {Coriat}, {Tzioumis}, {Read},
  {Knigge}, {Gandhi}, {Pretorius}, \& {Ag{\'\i}s-Gonz{\'a}lez}}]{Miniutti2019}
{Miniutti}, G., {Saxton}, R.~D., {Giustini}, M., {et~al.} 2019, \nat, 573, 381,
  \dodoi{10.1038/s41586-019-1556-x}

\bibitem[{{Mockler} {et~al.}(2022){Mockler}, {Twum}, {Auchettl}, {Dodd},
  {French}, {Law-Smith}, \& {Ramirez-Ruiz}}]{Mockler2022ApJ}
{Mockler}, B., {Twum}, A.~A., {Auchettl}, K., {et~al.} 2022, \apj, 924, 70,
  \dodoi{10.3847/1538-4357/ac35d5}

\bibitem[{{Nicholl} {et~al.}(2024){Nicholl}, {Pasham}, {Mummery}, {Guolo},
  {Gendreau}, {Dewangan}, {Ferrara}, {Remillard}, {Bonnerot}, {Chakraborty},
  {Hajela}, {Dhillon}, {Gillan}, {Greenwood}, {Huber}, {Janiuk}, {Salvesen},
  {van Velzen}, {Aamer}, {Alexander}, {Angus}, {Arzoumanian}, {Auchettl},
  {Berger}, {de Boer}, {Cendes}, {Chambers}, {Chen}, {Chornock}, {Fulton},
  {Gao}, {Gillanders}, {Gomez}, {Gompertz}, {Fabian}, {Herman}, {Ingram},
  {Kara}, {Laskar}, {Lawrence}, {Lin}, {Lowe}, {Magnier}, {Margutti}, {McGee},
  {Minguez}, {Moore}, {Nathan}, {Oates}, {Patra}, {Ramsden}, {Ravi}, {Ridley},
  {Sheng}, {Smartt}, {Smith}, {Srivastav}, {Stein}, {Stevance}, {Turner},
  {Wainscoat}, {Weston}, {Wevers}, \& {Young}}]{Nicholl2024}
{Nicholl}, M., {Pasham}, D.~R., {Mummery}, A., {et~al.} 2024, \nat, 634, 804,
  \dodoi{10.1038/s41586-024-08023-6}

\bibitem[{{Offner} {et~al.}(2023){Offner}, {Moe}, {Kratter}, {Sadavoy},
  {Jensen}, \& {Tobin}}]{Offner2023}
{Offner}, S.~S.~R., {Moe}, M., {Kratter}, K.~M., {et~al.} 2023, in Astronomical
  Society of the Pacific Conference Series, Vol. 534, Protostars and Planets
  VII, ed. S.~{Inutsuka}, Y.~{Aikawa}, T.~{Muto}, K.~{Tomida}, \& M.~{Tamura},
  275, \dodoi{10.48550/arXiv.2203.10066}

\bibitem[{{Paxton} {et~al.}(2011){Paxton}, {Bildsten}, {Dotter}, {Herwig},
  {Lesaffre}, \& {Timmes}}]{Paxton2011}
{Paxton}, B., {Bildsten}, L., {Dotter}, A., {et~al.} 2011, \apjs, 192, 3,
  \dodoi{10.1088/0067-0049/192/1/3}

\bibitem[{{Paxton} {et~al.}(2013){Paxton}, {Cantiello}, {Arras}, {Bildsten},
  {Brown}, {Dotter}, {Mankovich}, {Montgomery}, {Stello}, {Timmes}, \&
  {Townsend}}]{Paxton2013}
{Paxton}, B., {Cantiello}, M., {Arras}, P., {et~al.} 2013, \apjs, 208, 4,
  \dodoi{10.1088/0067-0049/208/1/4}

\bibitem[{{Paxton} {et~al.}(2015){Paxton}, {Marchant}, {Schwab}, {Bauer},
  {Bildsten}, {Cantiello}, {Dessart}, {Farmer}, {Hu}, {Langer}, {Townsend},
  {Townsley}, \& {Timmes}}]{Paxton2015}
{Paxton}, B., {Marchant}, P., {Schwab}, J., {et~al.} 2015, \apjs, 220, 15,
  \dodoi{10.1088/0067-0049/220/1/15}

\bibitem[{{Paxton} {et~al.}(2018){Paxton}, {Schwab}, {Bauer}, {Bildsten},
  {Blinnikov}, {Duffell}, {Farmer}, {Goldberg}, {Marchant}, {Sorokina},
  {Thoul}, {Townsend}, \& {Timmes}}]{Paxton2018}
{Paxton}, B., {Schwab}, J., {Bauer}, E.~B., {et~al.} 2018, \apjs, 234, 34,
  \dodoi{10.3847/1538-4365/aaa5a8}

\bibitem[{{Paxton} {et~al.}(2019){Paxton}, {Smolec}, {Schwab}, {Gautschy},
  {Bildsten}, {Cantiello}, {Dotter}, {Farmer}, {Goldberg}, {Jermyn}, {Kanbur},
  {Marchant}, {Thoul}, {Townsend}, {Wolf}, {Zhang}, \& {Timmes}}]{Paxton2019}
{Paxton}, B., {Smolec}, R., {Schwab}, J., {et~al.} 2019, \apjs, 243, 10,
  \dodoi{10.3847/1538-4365/ab2241}

\bibitem[{Pejcha {et~al.}(2016{\natexlab{a}})Pejcha, Metzger, \&
  Tomida}]{pejcha_cool_2016}
Pejcha, O., Metzger, B.~D., \& Tomida, K. 2016{\natexlab{a}}, Monthly Notices
  of the Royal Astronomical Society, 455, 4351, \dodoi{10.1093/mnras/stv2592}

\bibitem[{Pejcha {et~al.}(2016{\natexlab{b}})Pejcha, Metzger, \&
  Tomida}]{pejcha_binary_2016}
---. 2016{\natexlab{b}}, Monthly Notices of the Royal Astronomical Society,
  461, 2527, \dodoi{10.1093/mnras/stw1481}

\bibitem[{Peters \& Mathews(1963)}]{Peters_1963}
Peters, P.~C., \& Mathews, J. 1963, Phys. Rev., 131, 435,
  \dodoi{10.1103/PhysRev.131.435}

\bibitem[{Pols(2011)}]{pols2011stellar}
Pols, O. 2011, Stellar Structure and Evolution (Astronomical Institute
  Utrecht).
\newblock \url{https://books.google.de/books?id=sawHtAEACAAJ}

\bibitem[{Quataert(2024)}]{Quataert-KITP-2024}
Quataert, E. 2024, {The Fates of Stars Orbiting too Close to Massive Black
  Holes},  The Kavli Institute for Theoretical Physics, \dodoi{10.26081/K6FB0B}

\bibitem[{{Rafikov}(2020)}]{Rafikov2020}
{Rafikov}, R.~R. 2020, \apjl, 905, L35, \dodoi{10.3847/2041-8213/abcebc}

\bibitem[{{Robson} {et~al.}(2019){Robson}, {Cornish}, \& {Liu}}]{Robson2019}
{Robson}, T., {Cornish}, N.~J., \& {Liu}, C. 2019, Classical and Quantum
  Gravity, 36, 105011, \dodoi{10.1088/1361-6382/ab1101}

\bibitem[{{Rom} \& {Sari}(2025)}]{Barak2025}
{Rom}, B., \& {Sari}, R. 2025, arXiv e-prints, arXiv:2502.13209.
\newblock \doarXiv{2502.13209}

\bibitem[{{Ruan} {et~al.}(2018){Ruan}, {Guo}, {Cai}, \& {Zhang}}]{Taiji2018}
{Ruan}, W.-H., {Guo}, Z.-K., {Cai}, R.-G., \& {Zhang}, Y.-Z. 2018, arXiv
  e-prints, arXiv:1807.09495, \dodoi{10.48550/arXiv.1807.09495}

\bibitem[{{Ryu} {et~al.}(2023){Ryu}, {Krolik}, \& {Piran}}]{Taeho2023}
{Ryu}, T., {Krolik}, J., \& {Piran}, T. 2023, \apjl, 946, L33,
  \dodoi{10.3847/2041-8213/acc390}

\bibitem[{{Ryu} {et~al.}(2020){Ryu}, {Krolik}, {Piran}, \& {Noble}}]{Taeho2020}
{Ryu}, T., {Krolik}, J., {Piran}, T., \& {Noble}, S.~C. 2020, \apj, 904, 98,
  \dodoi{10.3847/1538-4357/abb3cf}

\bibitem[{{Salpeter}(1955)}]{Salpeter1955}
{Salpeter}, E.~E. 1955, \apj, 121, 161, \dodoi{10.1086/145971}

\bibitem[{{Shakura} \& {Postnov}(1987)}]{Shakura+Postnov1987}
{Shakura}, N.~I., \& {Postnov}, K.~A. 1987, \aap, 183, L21

\bibitem[{Shu {et~al.}(1979)Shu, Lubow, \& Anderson}]{shu_structure_1979}
Shu, F.~H., Lubow, S.~H., \& Anderson, L. 1979, The Astrophysical Journal, 229,
  223, \dodoi{10.1086/156948}

\bibitem[{{Syer} {et~al.}(1991){Syer}, {Clarke}, \& {Rees}}]{Syer+1991}
{Syer}, D., {Clarke}, C.~J., \& {Rees}, M.~J. 1991, \mnras, 250, 505,
  \dodoi{10.1093/mnras/250.3.505}

\bibitem[{{Tagawa} \& {Haiman}(2023)}]{Tagawa2023}
{Tagawa}, H., \& {Haiman}, Z. 2023, \mnras, 526, 69,
  \dodoi{10.1093/mnras/stad2616}

\bibitem[{{Tauris} \& {van den Heuvel}(2006)}]{Tauris2006}
{Tauris}, T.~M., \& {van den Heuvel}, E.~P.~J. 2006, in Compact stellar X-ray
  sources, ed. W.~H.~G. {Lewin} \& M.~{van der Klis}, Vol.~39 (Cambridge
  Astrophysics Series), 623--665, \dodoi{10.48550/arXiv.astro-ph/0303456}

\bibitem[{{van Kerkwijk} {et~al.}(2010){van Kerkwijk}, {Rappaport}, {Breton},
  {Justham}, {Podsiadlowski}, \& {Han}}]{vanKerkwijk+2010}
{van Kerkwijk}, M.~H., {Rappaport}, S.~A., {Breton}, R.~P., {et~al.} 2010,
  \apj, 715, 51, \dodoi{10.1088/0004-637X/715/1/51}

\bibitem[{{Wang}(2024)}]{Wang2024}
{Wang}, D. 2024, \aap, 687, A295, \dodoi{10.1051/0004-6361/202449585}

\bibitem[{{Xian} {et~al.}(2021){Xian}, {Zhang}, {Dou}, {He}, \&
  {Shu}}]{Xian2021}
{Xian}, J., {Zhang}, F., {Dou}, L., {He}, J., \& {Shu}, X. 2021, \apjl, 921,
  L32, \dodoi{10.3847/2041-8213/ac31aa}

\bibitem[{{Yao} {et~al.}(2024){Yao}, {Quataert}, {Jiang}, {Lu}, \&
  {White}}]{Yao2024}
{Yao}, P.~Z., {Quataert}, E., {Jiang}, Y.-F., {Lu}, W., \& {White}, C.~J. 2024,
  arXiv e-prints, arXiv:2407.14578, \dodoi{10.48550/arXiv.2407.14578}

\bibitem[{{York} {et~al.}(2000){York}, {Adelman}, {Anderson}, {Anderson},
  {Annis}, {Bahcall}, {Bakken}, {Barkhouser}, {Bastian}, {Berman}, {Boroski},
  {Bracker}, {Briegel}, {Briggs}, {Brinkmann}, {Brunner}, {Burles}, {Carey},
  {Carr}, {Castander}, {Chen}, {Colestock}, {Connolly}, {Crocker}, {Csabai},
  {Czarapata}, {Davis}, {Doi}, {Dombeck}, {Eisenstein}, {Ellman}, {Elms},
  {Evans}, {Fan}, {Federwitz}, {Fiscelli}, {Friedman}, {Frieman}, {Fukugita},
  {Gillespie}, {Gunn}, {Gurbani}, {de Haas}, {Haldeman}, {Harris}, {Hayes},
  {Heckman}, {Hennessy}, {Hindsley}, {Holm}, {Holmgren}, {Huang}, {Hull},
  {Husby}, {Ichikawa}, {Ichikawa}, {Ivezi{\'c}}, {Kent}, {Kim}, {Kinney},
  {Klaene}, {Kleinman}, {Kleinman}, {Knapp}, {Korienek}, {Kron}, {Kunszt},
  {Lamb}, {Lee}, {Leger}, {Limmongkol}, {Lindenmeyer}, {Long}, {Loomis},
  {Loveday}, {Lucinio}, {Lupton}, {MacKinnon}, {Mannery}, {Mantsch}, {Margon},
  {McGehee}, {McKay}, {Meiksin}, {Merelli}, {Monet}, {Munn}, {Narayanan},
  {Nash}, {Neilsen}, {Neswold}, {Newberg}, {Nichol}, {Nicinski}, {Nonino},
  {Okada}, {Okamura}, {Ostriker}, {Owen}, {Pauls}, {Peoples}, {Peterson},
  {Petravick}, {Pier}, {Pope}, {Pordes}, {Prosapio}, {Rechenmacher}, {Quinn},
  {Richards}, {Richmond}, {Rivetta}, {Rockosi}, {Ruthmansdorfer}, {Sandford},
  {Schlegel}, {Schneider}, {Sekiguchi}, {Sergey}, {Shimasaku}, {Siegmund},
  {Smee}, {Smith}, {Snedden}, {Stone}, {Stoughton}, {Strauss}, {Stubbs},
  {SubbaRao}, {Szalay}, {Szapudi}, {Szokoly}, {Thakar}, {Tremonti}, {Tucker},
  {Uomoto}, {Vanden Berk}, {Vogeley}, {Waddell}, {Wang}, {Watanabe},
  {Weinberg}, {Yanny}, {Yasuda}, \& {SDSS Collaboration}}]{Galaxy_Density2000}
{York}, D.~G., {Adelman}, J., {Anderson}, John~E., J., {et~al.} 2000, \aj, 120,
  1579, \dodoi{10.1086/301513}

\bibitem[{{Zucker} {et~al.}(2006){Zucker}, {Alexander}, {Gillessen},
  {Eisenhauer}, \& {Genzel}}]{Zucker+2006}
{Zucker}, S., {Alexander}, T., {Gillessen}, S., {Eisenhauer}, F., \& {Genzel},
  R. 2006, \apjl, 639, L21, \dodoi{10.1086/501436}

\end{thebibliography}
\bibliographystyle{aasjournal}



\end{document}